\documentclass[twocolumn]{aastex701}
\usepackage{amsmath}
\newcommand{\Sc}{\mathrm{Sc}}
\newcommand{\Pe}{\mathrm{Pe}}
\newcommand{\Pm}{\mathrm{Pm}}
\newcommand{\BV}{Brunt-V\"ais\"al\"a}
\newcommand{\R}{\mathcal{R}}

\graphicspath{{figures/}} 

\begin{document}

\title{Bridging the Prandtl number gap: 3D simulations of thermohaline convection in astrophysical regimes}

\author[orcid=0000-0003-4323-2082,sname='Fraser',gname='Adrian']{Adrian E.~Fraser}
\altaffiliation{NSF AAPF fellow}
\affiliation{Department of Applied Mathematics, University of Colorado, Boulder, CO 80309, USA}
\email[show]{adrian.fraser@colorado.edu}

\begin{abstract}

Thermohaline convection (also known as fingering convection or thermohaline mixing) occurs in stellar radiation zones where a sufficient inversion of the mean molecular weight is present. This process mixes chemicals radially and occurs in a variety of stars, including near the luminosity bump on the red giant branch and potentially in polluted white dwarfs. 
Previous efforts to characterize this process using 3D simulations have been restricted to regimes far from actual stars: The Prandtl number $\Pr$---the ratio of the kinematic viscosity to thermal diffusivity---assumes values as low as $10^{-6}$ in stars, but 3D simulations have been restricted to $\Pr \gtrsim 10^{-2}$. 
For this reason, disagreements between observations and simulations are routinely dismissed as stemming from this $\Pr$ gap.
This letter bridges this gap and demonstrates that 3D simulations of thermohaline convection can be performed in stellar parameter regimes. Using a suite of simulations spanning previously studied regimes with $\Pr \gtrsim 10^{-2}$ down to $\Pr = 10^{-6}$, we demonstrate that the chemical mixing model of \citet{brown_chemical_2013} remains consistent with 3D simulations across both regimes. Therefore, tensions between this model and observations cannot be dismissed as resulting from a $\Pr$ gap, and must be resolved by considering additional physics.

\end{abstract}




\section{Introduction} \label{sec:intro}
The theory of stellar structure and evolution has advanced significantly in recent decades. One breakthrough has been the understanding that radiation zones---layers that 
do not undergo convection---host novel forms of turbulence that, over time, can significantly alter the rotation rate of a star's core \citep[e.g.,][]{spruit_dynamo_2002,fuller_slowing_2019,barker_GSF_2019,Tripathi_GSF} or a star's chemical composition \citep[e.g.,][]{salaris_chemical_2017,garaud_journey_2021}. 

One such form of turbulence that may drive chemical mixing in radiation zones is thermohaline convection (also known as fingering convection or thermohaline mixing), a form of double-diffusive convection (DDC; see review by \citealt{garaud_double-diffusive_2018}). Thermohaline convection can occur in radiation zones with stable thermal stratification (i.e., stable according to both the Schwarzschild and Ledoux criteria) and inversions of the local mean molecular weight $\mu$, such that $\mu$ increases with radius. Rapid thermal diffusion compared to the slow molecular diffusion of chemical inhomogeneities enables this diffusive instability to act despite even a strongly stable (i.e., sub-adiabatic) thermal gradient.

The possible occurrence and behavior of thermohaline convection in stars was first studied by \citet{ulrich_thermohaline_1972} and \citet{kippenhahn_thermohaline}. Later, \citet{CharbonnelZahn2007} 
identified thermohaline convection as a possible explanation for anomalous chemical mixing inferred from observations of red giant branch (RGB) stars by \citet{Gratton} (see also \citealt{shetrone_constraining_2019} for more recent observations, and \citealt{tayar_is_2022} for noteworthy concerns with this interpretation). 
Shortly after RGB stars evolve past a region on the HR diagram known as the luminosity bump, their surface chemical abundances change in a way that points to some form of chemical mixing in the radiation zones deep beneath their convective envelopes. Characterizing this 
``extra mixing" in RGB stars remains a key motivator for the study of thermohaline convection in stellar interiors, though other notable contexts across stellar astrophysics exist as well, including: 
\begin{itemize}
    \item polluted white dwarfs (WDs; \citealt{Deal_WD,Wachlin2017,Wachlin_WDs,bauer_increases_2018,bauer_polluted_2019,Dwomoh2023,Deal_WD_mass});
    \item magnetic field transport in crystallizing WDs \citep{fuentes_thermohaline_MLT_2023,Castro-Tapia_MWD}; 
    \item chemical signatures of planetary engulfment \citep{Sevilla_thermohaline,Behmard_thermohaline};
    \item accretor stars in massive binaries \citep{Renzo}; 
    \item carbon-rich, metal-poor stars \citep{stancliffe_carbon-enhanced_2007,stancliffe_depletion_2009,placco_carbon-enhanced_2014};
    \item and lithium-rich red giants \citep{denissenkov_enhanced_2024}.
\end{itemize}

In each case, a common strategy for understanding the role of thermohaline convection in these objects is to compare observations (when available) against the results of 1D stellar evolution calculations that account for thermohaline convection by using simplified mixing prescriptions. 
A variety of prescriptions (or ``models") for chemical mixing by thermohaline convection have been put forth \citep[see review by][]{garaud_double-diffusive_2018}. The earliest and most widely used were introduced by \citet{ulrich_thermohaline_1972} and \citet{kippenhahn_thermohaline} and are broadly equivalent to the traditional mixing-length theory of convection \citep{fuentes_thermohaline_MLT_2023}---their models are simple and straightforward to implement, but include a free parameter whose plausible values 
vary by nearly three orders of magnitude \citep[see Sec.~2.2 of][]{CantielloLanger2010}. 
The assumed value of this parameter significantly affects the level of agreement when comparing observations to 1D calculations, presenting a major problem for such comparisons\footnote{Note that, regardless of the value one chooses for this parameter, the predicted turbulent diffusivity as a function of stratification disagrees very clearly with 3D hydrodynamic simulations; see Fig.~3 of \citet{traxler_numerically_2011}.}: When observations contradict the outputs of these 1D calculations, it can either indicate that the free parameter needs re-calibration, or that additional processes beyond hydrodynamic thermohaline convection are needed to explain observations. 
Importantly, it is the largest of the plausible values of this parameter (corresponding to rapid mixing) that is required in order to bring RGB observations and 1D calculations into agreement \citep{CantielloLanger2010}.

More recently, mixing prescriptions have been put forth that are informed by first principles and calibrated against 3D simulations. In particular, \citet{brown_chemical_2013} (hereafter BGS13) introduced such a model, building off of similar work in the oceanographic regime by \citet{RadkoSmith2012} (see also \citealt{harrington_enhanced_2019} and \citealt{fraser_magnetized_2024} for models that include magnetic fields). 

Because the chemical fluxes predicted by the BGS13 model are in such good agreement with those measured in 3D simulations, this model can be taken as a reliable predictor of chemical mixing by hydrodynamic thermohaline convection in stellar interiors. However, two concerns are often raised. First, the predictions of this model are at odds with RGB observations \emph{if one assumes that hydrodynamic thermohaline convection is the primary source of extra mixing}. Specifically, the BGS13 model predicts relatively little mixing compared to the more rapid mixing demanded by observations \citep{Wachlin_RGBs}. Second, the simulations used to calibrate the BGS13 model (and, more broadly, all 3D simulations of thermohaline convection to date) were performed in regimes far removed from actual stars, as detailed in Sec.~\ref{sec:gap} below \citep[this gap applies to early 2D simulations and laboratory experiments as well, see][]{CantielloLanger2010}. Thus, while the BGS13 model reliably predicts chemical fluxes in regimes probed by existing 3D simulations, the concern is often raised that it may be unreliable when extrapolated to astrophysical regimes. 

Of these two concerns, the former may be addressed by considering the effects of magnetic fields \citep[see][]{harrington_enhanced_2019,fraser_magnetized_2024}. 
This letter addresses the latter concern.
We show that 3D simulations in astrophysically relevant regimes are attainable, presenting a suite of simulations extending from previously studied regimes to astrophysical regimes. Our key finding is that the BGS13 model remains just as accurate in these newly simulated regimes.


\section{Simulations to date and the Prandtl number gap} \label{sec:gap}
The dynamics of thermohaline convection are sensitive to the local gradients and molecular diffusivities of the fluid, which are captured by three key parameters. The first two, measuring diffusivities, are the Prandtl number $\Pr$ and diffusivity ratio $\tau$, which measure the ratios of kinematic viscosity $\nu$ and compositional diffusivity $\kappa_C$, respectively, to thermal diffusivity $\kappa_T$, i.e.,
\begin{equation}
    \Pr \equiv \frac{\nu}{\kappa_T}, \quad \text{and} \quad \tau \equiv \frac{\kappa_C}{\kappa_T}.
\end{equation}
The Schmidt number $\Sc \equiv \nu / \kappa_C = \Pr / \tau$ is often helpful when characterizing the dynamics as well. 
Here and throughout this letter, we take $C$ to represent the concentration of whatever solute is primarily responsible for the gradient in mean molecular weight $\mu$, so that $C$ is roughly analogous to $\mu$. 
The third parameter, the density ratio $R_0$, measures background gradients, and can be understood as the ratio of the stabilizing thermal stratification to the destabilizing compositional gradient. In terms of $N_T$, the {\BV} frequency based on thermal stratification alone (i.e., neglecting gradients in composition $C$) and $N_C$, the {\BV} frequency based on chemical composition alone (i.e., neglecting thermal stratification), $R_0$ may be written as (alternative definitions given in \citealt{cresswell_3d_2025})
\begin{equation}
    R_0 \equiv \frac{|N_T^2|}{|N_C^2|}.
\end{equation}
A layer is stable to convection by the Ledoux criterion if $R_0 \geq 1$, and is linearly unstable to thermohaline convection if $1 \leq R_0 \leq 1/\tau$. Note that \citet{traxler_numerically_2011} and BGS13 found it helpful to characterize the system using the reduced density ratio 
\begin{equation} \label{eq:r-def}
    r \equiv \frac{R_0 - 1}{\tau^{-1} - 1},
\end{equation}
instead of $R_0$. Thermohaline convection then occurs for $0 \leq r < 1$, with $r=0$ denoting the Ledoux threshold. In this work, we instead characterize background gradients using the Rayleigh ratio \citep{xie_reduced_2017}
\begin{equation} \label{eq:R-def}
    \R \equiv \frac{1}{R_0 \tau},
\end{equation}
which is approximately $1/r$ when $\tau \ll 1$ and $R_0 \gg 1$. In terms of $\R$, instability to convection according to the Ledoux criterion corresponds to $\R > 1/\tau$, and instability to thermohaline convection corresponds to $1/\tau > \R > 1$. Thus, $\varepsilon \equiv \R - 1$ becomes a helpful measure of supercriticality.

Both $\Pr$ and $\tau$ are typically $10^{-2}$ or larger in 3D simulations to date, while values of $10^{-6}$ or smaller are expected in the regions of interest in RGB stars\footnote{Note that 2D simulations in regimes relevant to RGB stars have been studied previously \citep{Denissenkov}, but it was later shown that 2D and 3D simulations are quite different in this regime \citep{garaud_2d_2015}.}. 
While BGS13 showed their model reliably predicts chemical fluxes across simulations that vary $\Pr$ and $\tau$ by more than an order of magnitude (from $1/3$ to $10^{-2}$), it is often pointed out---especially when the implications of such models introduce tension between stellar evolution calculations and observations---that trends might change across the large gap separating these values from those found in stars \citep{CantielloLanger2010,Angelou_thermohaline_2012,Koester2015,henkel_thermohaline_2017}. That is, perhaps the mixing by thermohaline convection is well-predicted by BGS13's model in numerically accessible regimes, but the dynamics somehow change when $\Pr$, $\tau \ll 10^{-2}$ in a way that the model does not capture. The uncomfortable alternative, for which we advocate in this letter, is that there is some physics missing (like magnetic fields) that is not adequately captured by existing stellar evolution calculations.

\begin{figure}
    \centering
    \includegraphics[width=\linewidth]{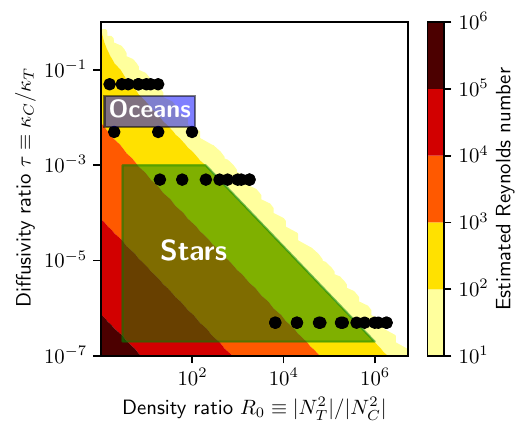}
    \caption{Different regimes in which thermohaline convection occurs (shaded regions) are compared against simulations presented in this work (black points) in terms of the 
    diffusivity ratio $\tau$ (characterizing the thermal diffusivity) and the density ratio $R_0$ (characterizing the stratification). Filled contours show the estimated Reynolds number of the flows (detailed in main text). 
    Thermohaline convection in oceans typically occurs with $\tau \sim 10^{-2}$; in stars, the possible values of $\tau$ vary dramatically, but are generally much smaller. 
    Previous 3D simulations have explored $\tau \gtrsim 10^{-2}$ extensively, while the present work considers simulations with much smaller $\tau$.
    }
    \label{fig:regimes}
\end{figure}

Such gaps between numerically accessible and astrophysically relevant regimes are ubiquitous whenever 3D direct numerical simulations (DNS) are used to study astrophysical systems. The solar convection zone, for instance, features dynamics with a Reynolds number (the ratio of inertial to viscous forces) of $10^{12}$--$10^{13}$ \citep{Ossendrijver,Matilsky}, implying an extreme range of scales between the largest and smallest eddies in the system. Simulating such flows via DNS requires far more grid points than are feasible on any modern or upcoming supercomputer. 
Unlike stellar convection, however, turbulence in radiation zones does not always feature such extreme Reynolds numbers. 
Figure \ref{fig:regimes} shows the estimated Reynolds number of hydrodynamic thermohaline convection as a function of $\tau$ (fixing $\Sc = 2$) and $R_0$. 
The Reynolds number is estimated taking the vertical velocity from the BGS13 model as the characteristic velocity and $100d$ as the characteristic length scale, where $d$ is the width of a typical finger (see Sec.~\ref{sec:setup} below), on the grounds that the most energetic eddies in 3D simulations of thermohaline convection are typically $10$--$100$ times larger than a finger width. 
Black points correspond to 3D simulations presented in this letter. The blue and green shaded regions indicate (very roughly) the parameters relevant to thermohaline convection in oceans and stars, respectively---in both cases, the horizontal extent of the shaded region is arbitrary (all values of $R_0$ are plausible, though see \citealt{fraser_characterizing_2022} for discussion of relevant values in RGB stars). While $\tau \sim 10^{-6}$--$10^{-7}$ is most relevant to the RGB case, larger values of $\tau$ may be found in WDs \citep{garaud_excitation_2015,bauer_increases_2018}. 
Importantly, while the Reynolds number might reach values as high as $10^6$ in stellar thermohaline convection, astrophysically relevant conditions exist where the Reynolds number may be $10^3$ or smaller, which is entirely accessible with modern computing capabilities.

\section{Simulation setup and nondimensionalization} \label{sec:setup}


Typical 3D simulations of thermohaline convection in stellar contexts are often carried out as follows (see review by \citealt{garaud_double-diffusive_2018}). First, cartesian domains are commonly used, justified by the small vertical extent of typical eddies compared to the radius of curvature (for simulations in spherical domains, see \citealt{guervilly_fingering_2022,tassin_fingering_2024}). 
Next, the Boussinesq approximation---which filters sound waves that travel much faster than the dynamics of interest---is justified on the grounds that the vertical extent is also small compared to the pressure scale height, and that the flow speed is always much slower than the local sound speed \citep{spiegel_boussinesq_1960}. 
Finally, it is assumed that the thermal and compositional fluxes driven by thermohaline convection are sufficiently small 
that we can impose fixed gradients that do not change with time and study fluctuations atop these gradients 
(\citealt{cresswell_3d_2025} and \citealt{zemskova_fingering_2014} studied astrophysically motivated setups 
that relaxed this assumption, and both found dynamics consistent with periodic models). 
In this work, we also neglect magnetic fields and rotation, though such effects have been considered elsewhere \citep{sengupta_effect_2018,harrington_enhanced_2019,fraser_magnetized_2024}. 


Under these assumptions, the dynamics are governed by a simple set of equations, which we express nondimensionally in terms of the following units \citep[following][]{xie_reduced_2017,fraser_helical_2025}: 
As our unit of length we take the characteristic finger width $d = (\kappa_T \nu/N_T^2)^{1/4}$; 
our unit of time is the compositional diffusion time across $d$ (as in \citealt{Prat_smallPr}); 
the unit temperature fluctuation is taken as the potential temperature difference (i.e., the subadiabaticity $\nabla - \nabla_\mathrm{ad}$) of the background gradient across a height $d$ multiplied by $\tau$ (consistent with the expectation of low-P\'eclet-number dynamics; see \citealt{Lignieres_LPN,Prat_smallPr,garaud_journey_2021,fraser_helical_2025}); and the unit compositional fluctuation is the compositional difference of the background across a height $d$ 
divided by $\R$. 

The resulting nondimensionalized equations are (dimensional form given in \citealt{garaud_double-diffusive_2018}):
\begin{equation} \label{eq:dimless-momB}
    \frac{1}{\Sc} \left( \frac{\partial \tilde{\mathbf{u}}}{\partial \tilde{t}} + \tilde{\mathbf{u}} \cdot \tilde{\nabla} \tilde{\mathbf{u}} \right) = - \tilde{\nabla} \tilde{p} + \tilde{\nabla}^2 \tilde{\mathbf{u}} + \left( \tilde{T} - \tilde{C} \right) \mathbf{e}_z,
\end{equation}
\begin{equation} \label{eq:dimless-TB}
    \tau \left( \frac{\partial \tilde{T}}{\partial \tilde{t}} + \tilde{\mathbf{u}} \cdot \tilde{\nabla} \tilde{T} \right) + \tilde{u}_z = \tilde{\nabla}^2 \tilde{T},
\end{equation}
\begin{equation} \label{eq:dimless-CB}
    \frac{\partial \tilde{C}}{\partial \tilde{t}} + \tilde{\mathbf{u}} \cdot \tilde{\nabla} \tilde{C} + \R \tilde{u}_z = \tilde{\nabla}^2 \tilde{C},
\end{equation}
and
\begin{equation} \label{eq:dimless-incompressibleB}
    \tilde{\nabla} \cdot \tilde{\mathbf{u}} = 0,
\end{equation}
where tildes denote nondimensional variables and derivatives, $\tilde{\mathbf{u}} = (\tilde u_x, \tilde u_y, \tilde u_z)$ is the velocity field, $\tilde{p}$ is the pressure, $\tilde T$ and $\tilde C$ are the thermal and compositional fluctuations (the latter analogous to mean molecular weight $\mu$) about the imposed linear background, and $\mathbf{e}_z$ is the unit vector in the $z$ direction. 
Note that our nondimensionalization differs from one commonly used throughout the DDC literature \citep[see][and references therein]{garaud_double-diffusive_2018}, though it is straightforward to convert between the two systems: let $\hat{f}$ represent some quantity $f$ that is nondimensionalized according to the units described in \citet{garaud_double-diffusive_2018}, then $\hat{\mathbf{u}} = \tau \tilde{\mathbf{u}}$, $\hat{T} = \tau \tilde T$, $\hat C = \tau \tilde C$, $\hat{\mathbf{x}} = \tilde{\mathbf{x}}$, and $\hat t = \tau^{-1} \tilde t$. The units employed here are thus equivalent to the traditional ones (as they must be), but they offer some convenience for studies that compare different values of $\Pr$ and $\tau$, as explained in Appendix \ref{sec:appendixA}. 



We solve Eqs.~\eqref{eq:dimless-momB}--\eqref{eq:dimless-incompressibleB} using the pseudospectral method as implemented in Dedalus v2 \citep{Dedalus2}, using periodic boundary conditions and expanding in Fourier bases along each axis. 
Domain sizes are set to integer multiples of the fastest-growing wavelength of the linear instability: We take $(L_x, L_y, L_z) = (8, 4, 16) \times 2\pi/\tilde{k}_{opt}$ (with $\tilde{k}_{opt}$ the fastest-growing wavenumber for a given set of parameters; note \citealt{garaud_2d_2015} showed that horizontally anisotropic domains with $L_x \neq L_y$ generally do not constrain the dynamics in this regime) except when simulating $\R \leq 1.1$, where we find twice the domain height is necessary to capture large-scale flows \citep{fraser_helical_2025}. 
The resolution required (as confirmed by inspecting for Gibbs ringing and checking that $\tilde{F}_C$ changes negligibly upon increasing the resolution along each axis) increases with $\R$: Eight Fourier modes per $2 \pi / \tilde{k}_{opt}$ in each direction suffices for $\R \leq 2$, sixteen for $\R \leq 31$, 32 for $\R \leq 101$, and 64 for our $\R = 301$ cases. We dealias using the standard $3/2$-rule, such that nonlinearities are evaluated on a grid with $3/2$ times this resolution. As in \citet{fraser_helical_2025}, we use a semi-implicit, 2nd-order Adams-Bashforth/backwards-difference scheme (Eq.~[2.8] of \citealt{Wang_timesteppers_2008}) with an advective CFL with safety factor of 0.3. We treat nonlinearities explicitly and all other terms implicitly---an important decision that we discuss further in Appendix  \ref{sec:appendixA}. 


\section{Results} \label{sec:results}

\begin{figure}
    \centering
    \includegraphics[width=\linewidth]{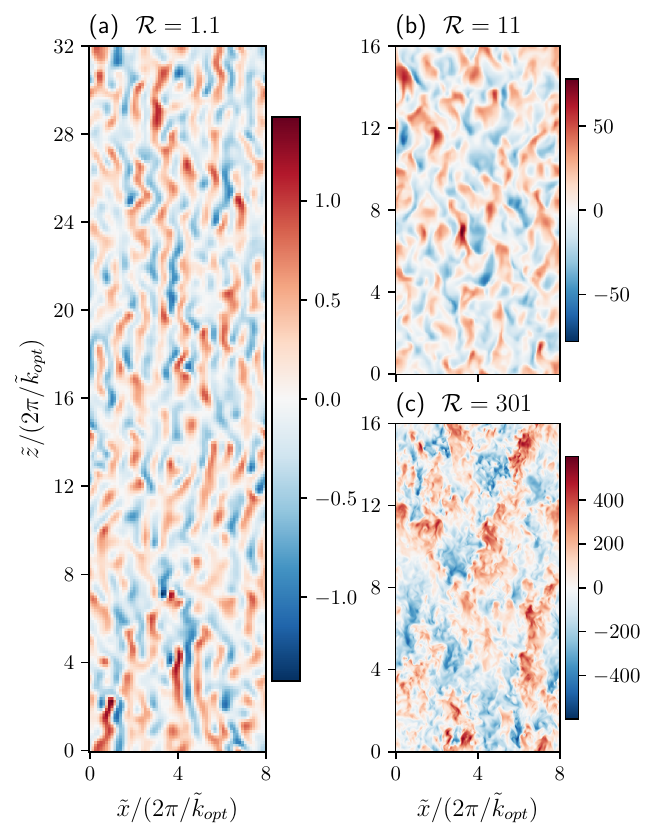}
    \caption{Snapshots of the vertical velocity $\tilde{u}_z$ at $\tilde{y} = 0$ in the saturated state for simulations with $\Pr = 10^{-6}$, $\tau = \Pr/2$, and (a) $\R = 1.1$, (b) $\R = 11$, and (c) $\R = 301$ (or $\varepsilon = 0.1$, $10$, and $300$). As $\R$ increases, velocity fluctuations increase and develop finer structures.
    }
    \label{fig:snapshots}
\end{figure}

We perform a suite of simulations varying $\R$, $\tau$, and $\Pr$ with fixed $\Sc = 2$. We perform some scans varying $\Pr$ and $\tau$ at fixed $r$ and $\Sc$, and others varying $\Pr$ and $\tau$ at fixed $\R$ and $\Sc$. We further supplement these scans with simulations of the $\tau \to 0$ equations (which replace Eq.~\eqref{eq:dimless-TB} with $\tilde{u}_z = \tilde{\nabla}^2 \tilde{T}$, leaving Eqs.~\eqref{eq:dimless-momB}, \eqref{eq:dimless-CB}, and \eqref{eq:dimless-incompressibleB} unchanged; c.f.~\citealt{Prat_smallPr}) at various values of $\R$. The resulting equations were studied in 2D by \citet{xie_reduced_2017,xie_jet_2019}, who referred to them as the modified Rayleigh-B\'enard convection system\footnote{Related systems, where $\Pr$ is fixed as $\tau \to 0$ and which may be relevant to WDs where $\Pr \gg \tau$ \citep{garaud_excitation_2015}, 
were also considered in 2D by \citet{xie_reduced_2017} and in 3D by \citet{fraser_helical_2025}}.

Figure \ref{fig:snapshots} shows snapshots of $\tilde{u}_z$ from simulations with $\Pr = 10^{-6}$, $\tau = 5 \times 10^{-7}$, and (a) $\R = 1.1$, (b) $\R = 11$, and (c) $\R = 301$. As the Rayleigh ratio $\R$ increases, the flow amplitudes increase and turbulent motions occupy a broader range of spatial scales. Simulations at these values of $\R$ but different values of $\Pr$ and $\tau$ (not shown) exhibit essentially identical flow structures and amplitudes as at $\Pr = 10^{-6}$ and $\tau = 5 \times 10^{-7}$. We note that simulations with the same $\R$ are much more comparable to one another than simulations with the same $r$. We find the helical mean flows explored in \citet{fraser_helical_2025} for $\R \lesssim 1.1$ for all values of $\Pr$ considered, but a thorough investigation of them is beyond the scope of this paper.

\begin{figure*}
    \centering
    \includegraphics[width=\linewidth]{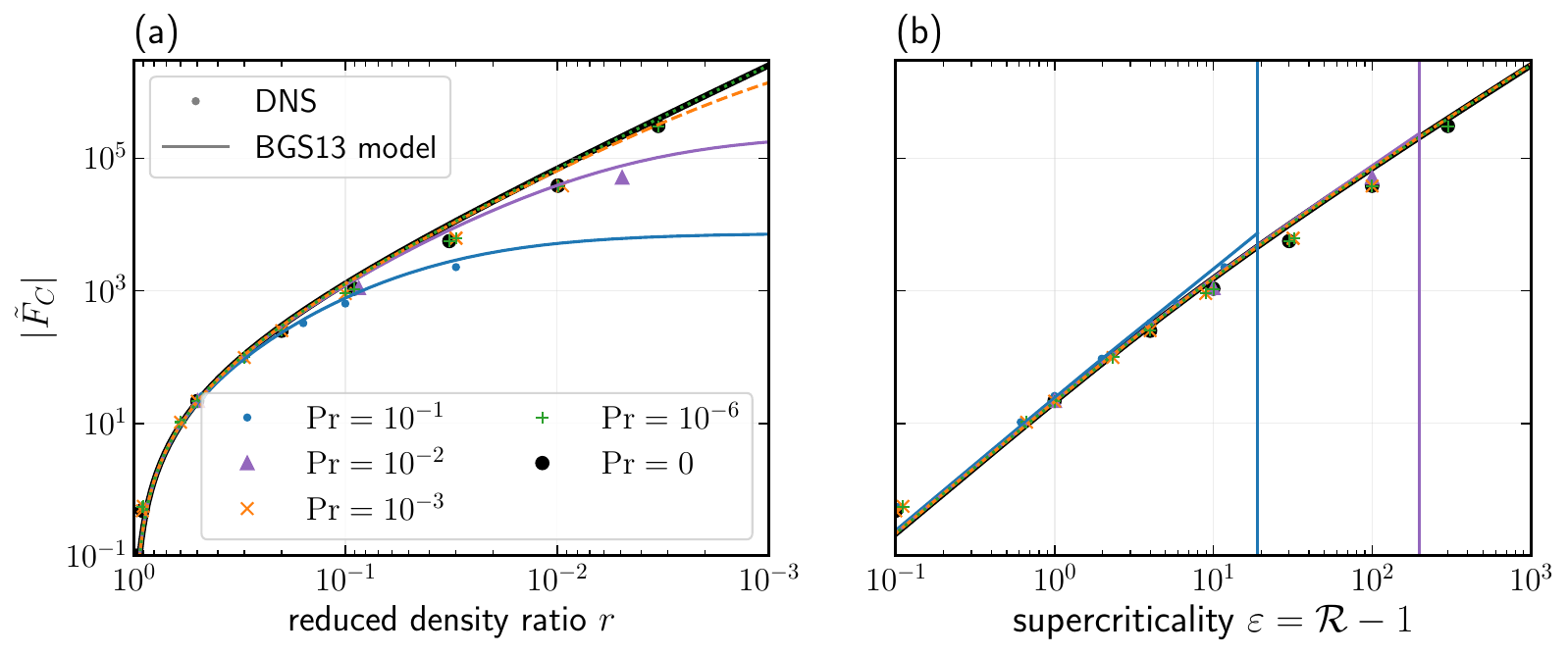}
    \caption{Turbulent compositional flux $|\tilde{F}_C|$ is shown across a range of parameters against (a) the reduced density ratio [see Eq.~\eqref{eq:r-def}] and (b) supercriticality $\varepsilon = \R - 1$. Points represent results from 3D simulations while curves show predictions from the BGS13 model, with color corresponding to the value of $\Pr$; throughout, we fix $\tau = \Pr/2$ ($\Sc = 2$). Vertical lines in panel (b) indicate the Ledoux threshold for convective instability for $\tau = 0.05$ (blue) and $\tau = 0.005$ (purple). Across all values of $\Pr$ considered, the BGS13 model remains in good agreement with simulations. Additionally, the data collapses more uniformly when plotted against $\varepsilon$ than it does when plotted against $r$, indicating the former may be a more practical control parameter when studying thermohaline convection at small $\Pr$ and $\tau$. 
    }
    \label{fig:DNS_scan}
\end{figure*}

Our key findings are shown in Fig.~\ref{fig:DNS_scan}, which shows the turbulent vertical compositional flux measured in these units, $|\tilde{F}_C| \equiv | \langle \tilde{u}_z \tilde{C} \rangle |$ (where $\langle \cdot \rangle$ denotes an average over the domain and over time throughout the saturated state; note this quantity is related to the compositional Nusselt number by $\mathrm{Nu}_C = 1 - \tilde{F}_C / \R$), as a function of (a) the reduced density ratio $r$ and (b) supercriticality $\varepsilon = 1 - \R$ for different values of $\Pr$ (different colors), with $\tau = \Pr/2$. Points show data extracted from simulations 
and curves show predictions from the BGS13 model. 
Vertical colored lines denote the Ledoux threshold ($R_0 = 1$, $\R = 1/\tau$) for the respective values of $\tau$.

The most important feature of Fig.~\ref{fig:DNS_scan} is that the BGS13 model retains the same accuracy at $\Pr = 10^{-6}$ as at larger values of $\Pr$. We find no evidence for any dynamics in the $\Pr = 10^{-6}$ regime that are not already present at larger $\Pr$, and thus tensions between the BGS13 model and observations cannot be ascribed to a gap in $\Pr$.

Additionally, curves and points corresponding to different $\Pr$ collapse almost uniformly in panel (b), except near the convective threshold (vertical lines) for a given $\tau$. Compared to the smaller degree of collapse in panel (a), this demonstrates that $\R$ (or $\varepsilon$) may be a more practical control parameter than $r$ in the regime of small $\Pr$ and small $\tau$. 

Finally, the $\Pr = 0$ and $\Pr = 10^{-6}$ cases nearly overlap. This suggests that the $\Pr = 0$ system can be relied upon for future studies of hydrodynamic thermohaline convection in these regimes. This is noteworthy because $\Pr = 0$ simulations are significantly cheaper: Replacing Eq.~\eqref{eq:dimless-TB} with $\tilde{u}_z = \tilde{\nabla}^2 \tilde{T}$ means fewer Fourier transforms must be computed at each step, the matrices used for implicit timestepping are lower-rank, and $\tilde{u}_z$ no longer needs to be stored in memory. Note that the reduced $\Pr = 0$ system is unlikely to be representative of the full dynamics in any regime where the P\'eclet number (the product of the Reynolds number multiplied and $\Pr$) exceeds unity, or where finite-P\'eclet-number effects---such as staircases or collectively-excited internal gravity waves \citep{traxler_dynamics_2011,stellmach_dynamics_2011}---are anticipated. While these scenarios are generally not expected in astrophysical regimes for the hydrodynamic case \citep{garaud_excitation_2015}, they may occur in the presence of a sufficiently strong magnetic field \citep{fraser_magnetized_2024}.

\section{Conclusions} \label{sec:summary}
We have presented 3D hydrodynamic simulations of thermohaline convection across a range of $\Pr$ and $\tau$, from the comparatively large values studied in existing work ($\Pr$, $\tau \gtrsim 10^{-2}$) to extremely small values ($\Pr = 10^{-6}$, $\tau = \Pr/2$) directly relevant to stellar interiors. 
Here, we discuss implications for stellar interiors. A broader discussion of direct numerical simulations in diffusive regimes for thermohaline convection and for other systems is provided in Appendix \ref{sec:appendixA}.


Across our large suite of simulations, we have compared the resulting vertical flux of chemical composition against the BGS13 model \citep{brown_chemical_2013}, which predicts chemical flux as a function of diffusivities and background gradients. We find that the BGS13 model---which had previously only been compared against 3D simulations in the $\Pr$, $\tau \gtrsim 10^{-2}$ regime---remains as accurate at $\Pr = 10^{-6}$ as at larger $\Pr$, with no need to re-calibrate its free parameters. Thus, the BGS13 model should be taken as a reliable predictor of radial chemical mixing by thermohaline convection in stars when rotation \citep{sengupta_effect_2018} and magnetic fields \citep{harrington_enhanced_2019,fraser_magnetized_2024} are negligible. When 
the implications of models like BGS13's are at odds with observations 
or otherwise introduce tension, 
this tension should not be disregarded as merely an artifact of insufficiently small Prandtl numbers used in simulations. 
This conclusion is especially pertinent to two contexts where the $\Pr$ gap has been raised: extra mixing in RGB stars, and the inferred accretion rates in hot, polluted WDs. 


In the RGB case, observations point to much faster mixing than models like BGS13's predict \citep{CantielloLanger2010,Wachlin_RGBs}. These models are thus at odds with observations if \emph{hydrodynamic} thermohaline convection is presumed to be the dominant source of extra mixing. The $\Pr$ gap is regularly identified as a reason to disregard the implications of 3D simulations like those in BGS13 and \citet{traxler_numerically_2011}, with authors instead incorporating faster mixing into stellar evolution calculations by applying ad-hoc increases to the free parameter in Kippenhahn's model \citep[e.g.,][]{Angelou_thermohaline_2012}, or introducing altogether new models fit more directly against observations \citep[e.g.,][]{henkel_thermohaline_2017}. 

In the case of polluted WDs, stellar evolution calculations that account for thermohaline convection 
find that the dilution rate of accreted material is significantly enhanced by thermohaline convection \citep{Deal_WD,Wachlin2017}, and that the inferred accretion rate in hot, young polluted WDs must be larger than previously believed by orders of magnitude \citep{bauer_increases_2018,bauer_polluted_2019,Wachlin_WDs,Dwomoh2023}. This implication has introduced considerable tension with the large body of calculations that neglect thermohaline convection. As in the RGB case, authors point to the Prandtl number gap as one of multiple reasons why the implications of models like BGS13's should be disregarded in polluted WDs \citep[see, e.g.,][]{Koester2015,Farihi_thermohaline_2016}.

In both cases, the BGS13 model introduces tensions that are real and 
demand a critical reevaluation of our understanding of these systems, or the inclusion of additional physical effects.
However, this work shows that the now-closed Prandtl number gap is not a valid reason to disregard the BGS13 model, and that problematic assumptions are more likely to be found elsewhere, either by considering additional sources of mixing in the RGB context \citep[as in, e.g.,][]{blouin_wave-driven_2025}, or by considering how thermohaline convection is modified by additional physical effects. As demonstrated by \citet{harrington_enhanced_2019} and \citet{fraser_magnetized_2024} (hereafter HG19 and FRG24, respectively), 
accounting for magnetic fields is an especially promising avenue: HG19 and FRG24 showed that even relatively weak, uniform magnetic fields (on the order of 100G) enhance chemical mixing by thermohaline convection enough to overcome the gap between RGB observations and the BGS13 model (see Fig.~7 of \citealt{fraser_magnetized_2024}). 
The effects of more complex magnetic fields, and of magnetic fields combined with rotation, have yet to be considered, and may similarly prove useful in reconciling such models with expectations in the context of polluted WDs. 
Similarly, \citet{sengupta_effect_2018} showed that rotation can greatly enhance chemical mixing in some regimes. 
Future work, therefore, should aim to characterize how magnetic fields of different strengths and configurations affect thermohaline convection, possibly in combination with rotation and/or external forcing by convectively excited internal gravity waves.



\begin{acknowledgments}
I thank Ian Grooms, Valentin Skoutnev, Evan Bauer, Brad Hindman, Pascale Garaud, Matteo Cantiello, Loren Matilsky, and Rich Townsend for useful conversations. Rich Townsend additionally provided a very helpful script for calculating $\tilde{k}_{opt}$, for which I am grateful. I also thank Nick Featherstone and Brad Hindman for generous help with numerical resources, and Alex Fus for proofreading an advanced copy of this manuscript. This letter is inspired by the work of the late Keith Julien (and his many collaborators, especially Edgar Knobloch), whose invaluable mentorship I gratefully acknowledge, and whose memory will always be a blessing. 

This work was supported by a National Science Foundation (NSF) Astronomy and Astrophysics Postdoctoral Fellowship (AAPF), under award AST-2402142. Resources supporting this work were provided by the NASA High-End Computing (HEC) Program through the NASA Advanced Supercomputing (NAS) Division at Ames Research Center.
\end{acknowledgments}

\software{Dedalus v2 \citep{Dedalus2}, matplotlib \citep{Matplotlib}, NumPy \citep{numpy}, SciPy \citep{2020SciPy-NMeth}
          }


\clearpage

\appendix
\section{Simulations in extremely diffusive regimes} \label{sec:appendixA}
\subsection{Thermohaline convection at small \texorpdfstring{$\Pr$}{Pr} and \texorpdfstring{$\tau$}{tau}}
Despite longstanding interest in thermohaline convection in stars, 3D simulations of this process in the astrophysically relevant regime of $\Pr$, $\tau \sim 10^{-6}$ have not been presented in prior works, as simulations in this regime were thought to be beyond modern computing capabilities. At the same time, \citet{Prat_smallPr} and \citet{fraser_helical_2025} presented 3D simulations in the asymptotic regime of $\tau \to 0$ (the former with finite $\Sc$, the latter with finite $\Pr$), raising the question of why the $\Pr$, $\tau \sim 10^{-6}$ regime should be inaccessible when both larger and (asymptotically) smaller values of $\Pr$ and $\tau$ can be achieved in 3D simulations. Here, we have drawn inspiration from recent breakthroughs in rapidly rotating convection by \citet{kan_bridging_2025} and \citet{julien_rescaled_2025} (see summary by \citealt{stellmach_new_2025}). In that context, 3D simulations had probed moderate rotation rates as well as the asymptotic limit of infinitely rapid rotation \citep{julien_generalized_2006,julien_heat_2012}, but simulations were unable to access the regime of finite-but-rapid rotation. \citet{kan_bridging_2025} and \citet{julien_rescaled_2025} used insights gained from the asymptotic limits of their system to identify a scheme for simulating the finite-but-rapidly rotating regime. Here, we have taken a similar approach for thermohaline convection: We have taken the same nondimensionalization and timestepping scheme as was used by \citet{fraser_helical_2025} to simulate the $\tau \to 0$ limit, but consider finite (but extremely small) $\tau$ by re-introducing the terms that had vanished upon taking the $\tau \to 0$ limit. 

Compared to the numerical scheme used in previous work \citep[described in][]{traxler_dynamics_2011}, the only differences are our choice of time-stepping scheme and of nondimensionalization.

We use a second-order timestepping scheme, whereas \citet{traxler_dynamics_2011} and subsequent works used a third-order scheme, but the key difference is in which terms are treated explicitly versus implicitly. We treat nonlinearities explicitly and \emph{all other terms implicitly}. We emphasize that 
this includes treating the buoyancy force in Eq.~\eqref{eq:dimless-momB} and the linear advection terms in Eqs.~\eqref{eq:dimless-TB} and \eqref{eq:dimless-CB} implicitly, which is in contrast with \citet{traxler_dynamics_2011} and subsequent works, which treat these terms explicitly. 
Either choice is valid, but treating these terms explicitly introduces numerical instabilities if the timestep is not sufficiently small compared to the inverse of the buoyancy frequency. This is a very strict constraint compared to the timescale of instability in the limit of small $\tau$ for all but the largest $\R$: In units of the compositional diffusion time, the buoyancy frequency is roughly $\tilde{\omega}_{buoy} \approx \sqrt{\Sc ( \tau^{-1} - \R)}$, 
while the growth rate in this regime scales as simply $\R$ to some positive power with an $\Sc$-dependent coefficient. Thus, treating these terms explicitly demands timesteps that are extremely short compared to the instability e-folding time (which is also roughly the typical eddy turnover time, as implied by the parasite models of \citealt{RadkoSmith2012,brown_chemical_2013}). 
Treating these terms implicitly permits larger timesteps that are numerically stable at the expense of treating internal gravity waves (IGWs) inaccurately by artificially damping them. However, we expect IGWs to be absent in this regime anyway. First, IGWs are already critically damped in generic low-P\'eclet flows, as demonstrated by \citet{Lignieres_LPN}. 
Second, it was carefully shown by \citet{garaud_excitation_2015} that IGWs are not excited in this regime through secondary instability mechanisms. As verification, we have checked that repeating some of the $\Pr = 0.1$ simulations with a timestep bounded by a fraction of $\tilde{\omega}_{buoy}$ leads to the same dynamics.
Thus, treating these terms implicitly is justified in this regime and enables simulations with timesteps that are only constrained by the advective CFL condition. 
We expect this timestepping scheme to be inappropriate for moderate or large P\'eclet numbers $\Pe \gtrsim 1$, or where large-P\'eclet dynamics, such as collectively excited IGWs or spontaneous staircase formation, are expected at large scales \citep{radko_mechanism_2003,traxler_dynamics_2011,stellmach_dynamics_2011,garaud_excitation_2015}, as might occur in scenarios where magnetic fields enhance flow speeds \citep{harrington_enhanced_2019,fraser_magnetized_2024}.

While our choice of nondimensionalization is not a strict necessity for enabling simulations in these regimes, it is a practical choice. We find that the dynamical fields ($\tilde{\mathbf{u}}$, $\tilde T$, $\tilde C$) all remain $O(1)$ in these units as $\tau \to 0$, which is not true for the more widely used units \citep[as used in, e.g.,][]{garaud_double-diffusive_2018}. Thus, when initializing a simulation at low $\tau$ with noise that one intends to be small-amplitude, it is easy to inadvertently initialize with large-amplitude noise when using the standard units. For example, a simulation at $\Pr = 2\tau = 10^{-1}$ and $\R = 1.1$ initialized with noise in $\hat{C}$ of amplitude $10^{-4}$ (recall $\hat{C} = \tau \tilde{C}$, see Sec.~\ref{sec:setup}) will have a well-defined linear growth phase before saturation, while the same initial condition for a simulation with $\Pr = 2\tau = 10^{-6}$ corresponds to a very large-amplitude initial condition that may lead to finite-time blowup for the same resolution and timestep size. However, initializing simulations with noise of amplitude $\tilde{C} \sim 10^{-4}$ in both cases permits a well-defined linear growth phase.

\subsection{Simulating highly diffusive regimes in other systems}
In stratified and/or electrically conducting fluids, it is understood that studying turbulence via direct numerical simulation (DNS) is challenging whenever one diffusion coefficient is much larger than another, such as when the Prandtl number $\Pr$ or the magnetic Prandtl number $\Pm = \nu/\eta$ (the ratio of kinematic viscosity $\nu$ to magnetic diffusivity $\eta$) is very large or very small. 
Assuming fully-developed turbulence with a driving scale and inertial range at scales larger than the various dissipation scales (due to viscosity, thermal diffusion, or resistivity), large or small Prandtl numbers imply a vast separation of scales between the driving scale and the smallest of the dissipation scales. Thus, simulating these regimes requires many grid points. Similarly, one might expect that thermohaline convection simulations with $\tau \ll 1$ should be inaccessible, because $\tau \ll 1$ might imply a large separation between the the thermal and compositional diffusion scales.

It bears clarifying how we have managed to simulate thermohaline convection with $\Pr, \tau \ll 1$ despite the above intuition, and under what conditions this intuition is similarly misleading in other systems. In short: the above primarily holds when the Reynolds number and P\'eclet number (or Reynolds and magnetic Reynolds number, in the case of MHD) characterizing the large, energy-containing scales are both large, and---with caveats detailed below---does not hold if one of these numbers is small.

In thermohaline convection with $\Pr, \tau \ll 1$, the linear instability always operates on thermally diffusive scales\footnote{This can be shown by calculating the nominal P\'eclet number of the instability as $\lambda_{opt}/(\kappa_T k_{opt}^2)$ where $k_{opt}$ is the most-unstable horizontal wavenumber and $\lambda_{opt}$ is its growth rate; for $\Pr, \tau \ll 1$, this quantity is always small \citep{brown_chemical_2013}.}, 
and so the notion of a dissipationless driving scale must be abandoned. 
Furthermore, 
because the coefficients in Eq.~\eqref{eq:dimless-momB} include only $\Sc$ rather than $\Pr$ and/or $\tau$ separately, and because the velocity $\tilde{\mathbf{u}}$ remains $O(1)$ in these units as $\Pr, \tau \to 0$, it follows that the Reynolds number does not change as $\Pr, \tau \to 0$ for fixed $\Sc$ and $\R$ (consistent with Fig.~\ref{fig:regimes}), and so the separation between the energy injection scales and the viscous scales remains fixed as $\Pr, \tau \to 0$ for fixed $\Sc$ and $\R$. 
Therefore, barring the secondary excitation of large-scale flows \citep[as occurs in other parameter regimes, see, e.g.,][]{stellmach_dynamics_2011} or additional physics that might enhance flow speeds (such as magnetic fields), the separation between the largest and smallest relevant scales remains fixed no matter how small $\Pr$ and $\tau$ become, and the resolution requirement for DNS at a particular $\Sc$ and $\R$ does not increase as $\Pr, \tau \to 0$. 
This reasoning extends to other systems as well: If turbulence is driven by an instability that always lies in dissipative scales such that either the P\'eclet, Reynolds, or magnetic Reynolds number is small, and the limit of large or small $\Pr$ or $\Pm$ can be taken in such a way that none of the P\'eclet, Reynolds, or magnetic Reynolds numbers grows, then, absent any large-scale secondary dynamics (such as the generation of large-scale waves or coherent structures), extreme values of $\Pr$ and/or $\Pm$ are fully accessible to 3D DNS. The Tayler instability in stellar interiors is an example of a system where extreme values of $\Pr$ and/or $\Pm$ are of interest and where, depending on the parameter regime, the instability can be inherently viscous, diffusive, or resistive, as shown by \citet{Skoutnev_TI}. By this reasoning, simulations of the Tayler instability with extreme values of $\Pr$ and/or $\Pm$ should be accessible provided these limits do not generate large-scale structures (including dynamo-generated magnetic fields) on increasingly large scales.

\section{Data availability}
Datasets used to generate all figures above, as well as scripts used to perform simulations and generate figures from datasets are available on Zenodo \citep{fraser_2026_19187749}. A git repository containing the same files without the datasets can be found at \texttt{github.com/afraser3/Bridging\_the\_Pr\_gap}.



\clearpage

\bibliography{thermohaline}{}
\bibliographystyle{aasjournalv7}



\end{document}